\documentstyle[preprint,aps,psfig]{revtex} 
\def\Journal#1#2#3#4{{#1} {\bf #2}, #3 (#4)}

\def\PRL{\em Phys. Rev. Lett.}
\def\PRD{{Phys. Rev.} D}

\def\AA{{\em Astron. Astrophys.}}

\def\CQG{{Class. Quant. Grav.}}

\def\be{\begin{equation}}
\def\ee{\end{equation}}
\def\bea{\begin{eqnarray}}
\def\eea{\end{eqnarray}}

\tighten
\begin{document}

\renewcommand{\thefootnote}{\fnsymbol{footnote}} 
\def\scf{\setcounter{footnote}}

\hyphenation{coor-don-nees Ex-pli-cite-ment res-pec-tive-ment Le-gen-dre 
pro-pa-ga-tion va-ria-bles}
\title{On spontaneous scalarization  
\\ } 
\author{Marcelo Salgado\footnote{e-mail: marcelo@nuclecu.unam.mx}, 
Daniel Sudarsky \\ 
\small {\it Instituto de Ciencias Nucleares} \\ \small
{\it Universidad Nacional Aut\'onoma de M\'exico} \\ \small {\it 
Apdo. Postal 70-543 M\'exico 04510 D.F, M\'exico}\\ and \\
Ulises Nucamendi \\
\small {\it Departamento de F\'{\i}sica}\\
\small {\it Centro de Investigaci\'on y de Estudios Avanzados del I.P.N.}\\
\small {\it A. P. 14-741, M\'exico, D. F. 07000,  M\'exico.} }
\maketitle
\centerline {\bf Abstract}

We study in the physical frame the phenomenon of spontaneous scalarization
that occurs in scalar-tensor theories of gravity for compact objects. We
discuss the fact that the phenomenon occurs exactly in the regime where
the Newtonian analysis indicates it should not.  Finally we discuss the
way the phenomenon depends on the equation of state used to describe the
nuclear matter.  
 
 \vskip 1.5cm
\noindent PACS number(s): 04.50.+h, 97.60.Jd, 04.25.Dm

\newpage




\newpage

\section{Introduction}

\bigskip
In recent years, alternative field theories of gravity 
(scalar-tensor theories) have been analyzed in many contexts 
(see \cite{damour-1} for a review), specially as a way to study 
 possible deviations from general relativity and trough comparison  
with experimental results put bounds on the magnitude of such deviations.
 
The basic assumption of these theories is the existence of 
a single or multiple fundamental scalar fields which are 
perhaps the effective relics of a fundamental unified 
theory like superstrings, supergravity or Kaluza-Klein theory, 
and that might couple to gravity either minimally or non-minimally.

One of the most interesting effects occurring in a certain class of
scalar-tensor theories of gravity is the phenomenon of spontaneous
scalarization in neutron stars that was recently discovered by Damour and
Esposito-Far\`ese \cite{damour1,damour2}, which consists in the appearance
of a non-trivial configuration of a scalar field in the absence of sources
and with vanishing asymptotic value. 
 This phenomenon requires the presence of a scalar field nonminimally
coupled to gravity. This nonminimal coupling can be thought as a
field-dependent gravitational constant in the spirit of the Brans-Dicke
theory \cite{will,salmona,hillde,saenz}. 
 The occurrence of this highly nonlinear effect can be used to put very
 stringent limits on the deviations from general relativity, and in fact
using the binary pulsar, the limits for certain range of parameters seem
to be more severe than those that can be inferred from solar system
experiments \cite{damour2}.

One puzzling aspect of the phenomenon is that occurs exactly in the regime
where the Newtonian analysis indicates it should not,
  while on the other hand, no such phenomenon has been said to occur in
the
 regime where the Newtonian analysis indicates it should. We give a very
short description of these facts based on an energetic analysis and then
carry out a numerical study to
 examine its validity and explain why does the full fledge theory behave
in the opposite direction as compared to what is indicating by the
Newtonian intuition. To do this we have to perform the analysis in the
physical frame where a counterpart to the intuitive Newtonian ideas can be
found in contrast with the original analysis \cite{damour1,damour2}, which
was carried out in a conformal
 frame that, although yield simpler equations, corresponds to a description that
is less susceptible of comparison with the Newtonian analysis.

 Finally, and in view of the fact that the phenomenon when applied to the
binary pulsar
 is being used to put severe limits on the theory's parameters, we
investigate the degree to which the phenomenon depends on the equation of
state assumed for the description of the nuclear matter of neutron stars.

The paper is organized as follows. In section II we briefly review the
phenomenon of spontaneous scalarization. In section III we provide the
naive Newtonian energetic analysis of spontaneous scalarization. In
section IV we introduce the scalar-tensor theory we will analyze in
detail, derive the equations of motion for a spherically symmetric, static
space-time, and make some comments about the realistic equations of state
used in this work. In section V we discuss the boundary conditions for the
numerical integration, and define the global quantities to be computed and
which are of astrophysical interest. In section VI we provide the
numerical analysis and the explanation of the origin of spontaneous
scalarization on relativistic energetic grounds. Finally, in section VII
we discuss and summarize our results. 

\bigskip \section{Spontaneous scalarization } \bigskip As we mentioned,
the phenomenon of spontaneous scalarization was discovered by Damour and
Esposito-Far\`ese in scalar-tensor theories of gravity
\cite{damour1,damour2}, and further studied in the context of stability
analysis \cite{harada1,harada2}. However, we have to comment that Zagaluer
\cite{zagla} had previously argued for large deviations from general
relativity in neutron stars in a particular class of scalar-tensor
theories. Such results have been criticized in that seem spurious due to
an artificial definition of some parameters \cite{damour1}. 

In these theories, gravity is described both by the standard type (0,2) tensor 
field, and by one or more scalar fields. We concentrate for simplicity 
in the case of one scalar field, and the Lagrangian is then taken to be:
\be\label{stt}
{\cal L} = \left({ 1\over 16\pi G_0} \right)
\sqrt{-g} [f(\phi)R - z(\phi)(\nabla \phi)^2] 
+ {\cal L}_{\rm mat}(\psi,g_{\mu\nu}) \,\,\,\,.
\ee
Where  $G_0$ is the Newton's gravitational constant, 
$\psi$ denotes the matter fields, and $g_{\mu\nu}$ stands for 
the physical metric (Jordan metric). 
Using the conformal transformation to a non-physical metric 
$g^{*}_{\mu\nu}=f(\phi)g_{\mu\nu}$ 
(Einstein metric) and redefining a new scalar field,
\be
\varphi = \int \left[ {3\over 4} \frac{1}{f^2(\phi)} 
(\frac{\partial f(\phi)}{\partial \phi})^2 
+ {1\over 2} \frac{z(\phi)}{f(\phi)} \right]^{1/2} d\phi \,\,\,\,,
\ee
this Lagrangian can be brought in the form
\be
{\cal L} = \left({ 1\over 16\pi G_0} \right)
\sqrt{-g_{*}} [R_{*} - 2(\nabla_{*} \varphi)^2] 
+ {\cal L}_{\rm mat}(\psi,f^{-1}(\varphi)g^*_{\mu\nu}) \,\,\,\,.
\ee
Here the tensorial operations are performed by using the
non-physical metric $g^*_{\mu\nu}$.
The reason given in \cite{damour1,damour2} for writing the field 
equations in terms of the non-physical metric rather than 
the physical one is the mathematical simplicity related to 
the fact that the non-physical gravitational variables, $g^*_{\mu\nu}$ and 
$\varphi$, are minimally coupled. Many authors have also employed that frame to 
analyze neutron-star equilibrium configurations in other related 
contexts \cite{harada1,harada2,novak,japon}.

The field equations derived of the above Lagrangian are
\be
G^*_{\mu\nu} = 2\nabla^{*}_{\mu}\varphi\nabla^{*}_{\nu}\varphi
- g^*_{\mu\nu}(\nabla_{*} \varphi)^2 
+ 8\pi G_0 f^{-1}(\varphi) T_{\mu\nu} \,\,\,\,,
\ee
\be
\Box^{*}\varphi = 2\pi G_0 f^{-2}
(\partial \ln f/\partial\varphi) 
T,
\ee
\be
\nabla_{\mu} T^{\mu\nu} = 0 \,\,\,\,,
\ee
where $T^{\mu\nu}$ is the physical stress-energy tensor. 
For a particular class of scalar-tensor theories and using this frame 
(Einstein frame), the effects on 
neutron stars were investigated. 
Specifically, it was analyzed the equilibrium configurations 
of a spherically symmetric, static, isolated neutron stars 
using polytropic equations of state to model the nuclear matter 
\cite{damour1,harada1,harada2,novak,japon}, and also in 
axisymmetric slow rotating polytropes \cite{damour2}. 

 In particular, 
it was found that for the case of scalar-tensor theories with 
$\beta(\varphi_{0}=0)\leq-4$ 
[where $\beta(\varphi)=(-1/2)(\partial^2 \ln f/\partial\varphi^2)$], there 
existed a certain critical baryonic mass above which
 the star developed a non-zero scalar field 
$\varphi$,  even if $\varphi_{0}= 0$ (here $\varphi_{0}$ is the
 value at spatial infinity of the scalar field, i.e., the cosmological
value) \cite{damour1,damour2,harada1}. This occurs because such
configurations are energetically more favorable than the one having
$\varphi=0$ in all the space (both configurations having equal and
constant baryonic masses). This striking effect was baptized as
spontaneous scalarization in analogy with the spontaneous magnetization
arising in ferromagnets below the Curie temperature. These equilibrium
configurations lead to strong deviations from general relativity in the
dynamics of binary-pulsar systems \cite{damour2}. The fact that the
analysis was done in the non-physical frame does not allow us to
understand the interplay of the true contributions to the energy due to
the physical fields that are responsible of this phenomenon. We refer the
reader to Ref. \cite{dick} for a discussion of the problems associated
with this point. 

\bigskip  
\section{Newtonian energetic analysis}
\bigskip
One would like to understand the occurrence of
the phenomenon of spontaneous scalarization in  neutron stars on simple 
energetic grounds, much in the same way that one understands the occurrence of
 spontaneous magnetization on ferromagnetic materials by comparing 
the free energy of the 
states with no magnetization to the free energy of the states with 
magnetization arriving to the conclusion that for 
a sufficiently low temperature the free energy is minimized by the 
latter rather than the former.

 Let's try then to understand the phenomenon on these energetic terms using a
 Newtonian analysis of 
the contributions to the total energy of a given matter-scalar 
field configuration. 
First of all we recall that one expects the true
equilibrium configurations in this case to correspond to the minima
 of the energy
 within the space of configurations with fixed values of the other
 conserved quantities \cite{sudwald},
 in this case the only relevant conserved quantity is the total baryon 
number of the neutron star 
(for simplicity we are taking  the temperature to vanish). 
From this point of view one would think naively that since
 the net contribution of the scalar 
field to the energy is expected to be positive, the minimizing
 configuration should have 
 $\phi \equiv 0$.  A little thought then points to a caveat 
given the fact that in the class of scalar-tensor theory with 
$f(\phi)= (1 +16\pi\xi\phi^2)$, the effective gravitational 
constant depends on $\phi$, i.e., $G_{\rm eff} =  G_0/(1 +16\pi\xi\phi^2)$.
The point is that we expect the total energy of the configuration 
to be given by the  sum
of the mass energy of the neutrons (neglecting any 
thermal energy and those contributions 
arising from short range interactions), the energy of the scalar field,
  and the gravitational binding energy. Thus roughly we expect
\be
{\cal M} =  \int  \rho_{\rm bar} (\overline x) d^3 
\overline x  + \int  \rho_\phi (\overline x) d^3\overline x - 
G_{\rm eff} (\phi) \int   
\frac {\rho_{\rm bar} (\overline x) \rho_{\rm bar} (\overline y)}{|\overline x- 
\overline y|} d^3\overline x d^3\overline y +... \,\,\,\,,
\label{(2)}
\ee

where we took into account the dependence of the effective gravitational
 constant on $\phi$. Now we recall that we are expected to  minimize ${\cal M}$
 subject to the condition

\be
{\cal M}_{\rm bar} =  \int  \rho_{\rm bar} (\overline x)  d^3
\overline x = \qquad {\rm constant} \,\,\,\,.
\label{(3)}
\ee

 Thus the first term in Eq. (\ref{(2)}) is fixed, and the second is
expected to
 be positive, but it is conceivably that if the third integral is
sufficiently large (i.e., if the object is compact enough) then the
positive energy contribution of the scalar field can be more than
compensated by the change in the $G_{\rm eff}$ that would result in the
gravitational binding energy becoming substantially more negative, and
thus making the configurations with a non-vanishing scalar field
energetically preferred. This, of course, would require that
 $G_{\rm eff}$ be an increasing function of $\phi$, which in our specific
case corresponds
 to $\xi <0$. Thus we seem to have arrived to the conclusion that the
phenomenon of
 spontaneous scalarization is possible only if $G_{\rm eff}$ be an
increasing function
 of $\phi$. 
This is precisely the opposite of what is found in the relativistic 
analysis carried out in Refs. \cite{damour1,damour2}, and in the present paper.

We will examine next the phenomenon using the exact form of the theory in order 
to understand why did the naive Newtonian analysis fail so miserably.

\section{Formulation of the model}

We will consider a model of
a scalar field $\phi$ non-minimally coupled (NMC) to gravity. One of the
simplest models of this kind is obtained by considering the Lagrangian 
\be
{\cal L} = \left({ 1\over 16\pi G_0} + \xi \phi^2\right)
\sqrt{-g} R - \sqrt{-g} \left[ {1\over 2}(\nabla \phi)^2
+ V(\phi) \right] + {\cal L}_{\rm mat} \,\,\,\,\,.
\label{lag}
\ee
Here $\xi$ stands for the 
NMC constant, and $V(\phi)$ is a scalar potential. 
In this model the schematic matter Lagrangian ${\cal L}_{\rm mat}$ 
is intended to represent the ordinary matter contribution to the total 
Lagrangian.
 
As we mentioned, equation (\ref{lag}) shows that
the introduction of the coupling term is equivalent to consider 
an effective gravitational constant which explicitly depends
on the scalar field. We note that for $\xi$ positive, (the convention 
that we keep in all the paper),
the effective gravitational constant might only
 decrease with respect to the Newtonian value: 
\be
G_{\rm eff} ={G_0\over   1+ 16\pi G_0  \xi \phi^2}\ .
\label{geff}
\ee
The gravitational field equations following from the Lagrangian 
(\ref{lag}) can be written as
\be
R^{\mu\nu} - {1\over 2} g^{\mu\nu}R = 8\pi G_0 T^{\mu\nu}_{\rm eff}
\ee
where 
\bea\label{Teff}
T^{\mu\nu}_{\rm eff} &=& 
\frac{G_{\rm eff}}{G_0} \left(4\xi T^{\mu\nu}_\xi + T^{\mu\nu}_{\rm sf} +  
T^{\mu\nu}_{\rm mat}\right) \ ,\\
\label{feq}
T^{\mu\nu}_\xi &=& \nabla^\mu(\phi\nabla^\nu\phi) - g^{\mu\nu}
\nabla_\lambda (\phi \nabla^\lambda\phi) \ , 
\label{txi} \\
T^{\mu\nu}_{\rm sf} &=& \nabla^\mu\phi\nabla^\nu\phi
- g^{\mu\nu}\left[{1\over 2}  (\nabla \phi)^2 + V(\phi)\right] \ .
\label{tsf}
\eea

\noindent The energy-momentum tensor of matter $T^{\mu\nu}_{\rm mat}$ 
will be represented by a perfect fluid describing the cold catalyzed matter 
of a neutron star \cite{hww}:
\be
T^{\mu\nu}_{\rm mat}= (p + \rho ) U^\mu U^\nu + p g^{\mu\nu} \ ,
\label{tpf}
\ee
which possesses the symmetries of the spacetime. The scalar field 
will also be assumed to posses these symmetries. 

Finally, the equation of motion for the scalar field becomes
\be
\Box \phi + 2\xi \phi R = {\partial V(\phi)\over \partial \phi} \ .
\label{seq}
\ee

We will focus on a metric describing spherical and static space-times. 
Four our purposes it will be convenient to adopt 
the so-called {\it radial gauge} coordinates which are also 
{\it maximal slicing} coordinates (hereafter we refer these coordinates as to 
RGMS coordinates):
\be
ds^2 = -N^2(r) dt^2 + A^2(r)dr^2 + r^2d\theta^2
+ r^2\sin^2\theta d\varphi^2 \,\,\,\,\,.
\label{RGMS}
\ee
 
Our goal is to analyze solutions of the 
gravitational, matter, and scalar field equations 
 describing neutron-star models and the resulting 
space-time. Owing to the complexity of the resulting equations, we 
will perform a numerical analysis. To do so 
it will be helpful to adopt the following variables
\bea
\nu (r) &=& {\rm ln}[N(r)] \,\,\,\,\,,\\
\label{tildenu}
\tilde \nu (r) &=& \nu(r) - \nu(0)\,\,\,\,,\\
\label{AA}
A(r) &=& \left(1- \frac{2G_0m(r)}{r}\right)^{-1/2}\,\,\,\,\,.
\eea

The relevant Einstein equations take then the following form
\bea
\frac{\partial m}{\partial r}
&=&  4\pi r^2 E \,\,\,\,\,, \\
\frac{\partial \tilde \nu}{\partial r} &=&  
A^2 \left\{ \frac{G_0m}{r^2} + 4\pi r G_0 T^{r}_{{\rm eff}\,\,r} 
 \right\} \,\,\,\,. 
\eea
where
\be\label{E}
E= N^2 T^{tt}_{\rm eff} \,\,\,\,,
\ee
is the effective total energy density.

On the other hand, the Klein-Gordon equation can be written directly in terms 
of sources as follows:
\be
\Box \phi  =-16\pi\xi\phi G_0\left(E-S\right) + 
{\partial V(\phi)\over \partial \phi} \ .
\label{seq2}
\ee
where 
\be\label{S}
S= T^{i}_{{\rm eff}\,\,i}\,\,\,\,,
\ee 
is the trace of the spatial part of $T^{\mu\nu}_{\rm eff}$, which 
plays the role of an effective pressure.

In the RGMS coordinates this equation reads
\bea
\frac{\partial^2 \phi}{\partial r^2} &=& 
-\left[ \frac{2}{r} + \frac{\partial \tilde \nu}{\partial r} - 
\left(1-\frac{2 G_0 m}{r}\right)^{-1}
\left(4\pi G_0 r E -\frac{G_0 m}{r^2}\right) 
\right] \frac{\partial\phi}{\partial r} \nonumber \\
& & + \left(1-\frac{2G_0 m}{r}\right)^{-1} \left[
{\partial V(\phi)\over \partial \phi}
-16\pi\xi\phi G_0\left(E-S\right) \right] \,\,\,\,\,.
\eea

The matter equations are obtained from
\be
\nabla_\mu  T^{\mu \nu}_{\rm eff}=0 \,\,\,\,\,,
\ee
and straightforward manipulations with help of the Einstein equations 
show that the matter and the scalar-field energy-momentum tensors 
are conserved separately, leading thus to
\be
\nabla_\mu  T^{\mu \nu}_{\rm matt} =0\,\,\,\,\,.
\ee
This yields the equation of hydrostatic 
equilibrium which is similar to the one obtained in general relativity with 
no scalar fields.

For cold catalyzed matter (see below), it's customary to have the equations 
of state (EOS) parameterized by the baryon density $n(r)$, thus it is
 convenient to write the equation of hydrostatic equilibrium  
in terms of this:
\be\label{matter}
\frac{\partial n}{\partial r}= - \frac{ (\rho + p) n}{\gamma p} 
\frac{\partial \tilde \nu}{\partial r} \,\,\,\,,
\ee
where $\gamma:= \frac{d{\rm ln}[p]}{d{\rm ln}[\rho]}$ is the adiabatic index. 

It is clear from Eqs. (\ref{Teff}--\ref{tsf}), that the intermediary
variables $E$ and $S$ [see Eqs.(\ref{E}), (\ref{S})] involve second order
derivatives of the scalar field. 
 However, we can eliminate such a terms from the gravitational field
equations with the help of the Klein-Gordon equation, and obtain
``sources"  containing at most first order derivatives of the scalar
field. 
 We also introduce the following dimensionless quantities 
(where we restore the factors of $c$) 
\bea
\tilde r &:=& r\cdot \frac{c^2}{G_0 M_\odot}  \,\,\,\,,\\
\tilde m &:=& \frac{m}{M_\odot} \,\,\,\,,\\
\tilde n &:=& \frac{n}{.1 {\rm fm}^3}\,\,\,\,\,,\\
\label{rotilde}
\tilde \rho &:=& \rho\cdot \frac{G_0^3 M_\odot^2}{c^8}  \,\,\,\,,\\
\tilde p &:=& p\cdot \frac{G_0^3 M_\odot^2}{c^8}  \,\,\,\,,\\
\tilde \phi &:=& \phi\cdot \frac{\sqrt{G_0}}{c^2}  \,\,\,\,,\\
\tilde V(\tilde \phi) &:=& V(\tilde \phi)\cdot \frac{G_0^3 M_\odot^2 }{c^8}  
\,\,\,\,,\\
\tilde G_{\rm eff} &:=&\frac{1}{1+ 16\pi\xi\tilde\phi^2}\,\,\,\,,
\eea
then the final form of the equations to be solved numerically is :
\bea
\label{massu}
\partial_{\tilde r} \tilde m &=&  4\pi \tilde r^2 \tilde E \,\,\,\,\,,\\
\label{lapsefi}
\partial_{\tilde r}\tilde \nu  &=& \frac{A^2}{1+16\pi\xi 
\tilde r\phi (\partial_{\tilde r}\phi) \tilde G_{\rm eff} }
 \left\{ \frac{\tilde m}{\tilde r^2} + 4\pi \tilde r \tilde G_{\rm eff}\left[
\frac{1}{2A^2}(\partial_{\tilde r}\phi)^2 -\tilde 
V(\phi) + \tilde p -\frac{8\xi\phi\partial_{\tilde r}\phi}
{\tilde rA^2}\right]  \right\} \,\,\,\,, \\
\partial_{\tilde r\tilde r} \phi &=& 
-\left[ \frac{2}{\tilde r} + \partial_{\tilde r} \tilde \nu- 
\left(1-\frac{2\tilde m}{\tilde r}\right)^{-1}
\left(4\pi \tilde r \tilde E -\frac{\tilde m}{\tilde r^2}\right) 
\right] \partial_{\tilde r}\phi \nonumber \\
& & + \left(1-\frac{2\tilde m}{\tilde r}\right)^{-1} \left[
{\partial \tilde V(\phi)\over \partial \phi}
-16\pi\xi\phi \left(\tilde E-\tilde S\right) \right] \,\,\,\,\,,\\
\label{matterfi}
\partial_{\tilde r} \tilde n &=& - \frac{ (\tilde \rho + 
\tilde p) \tilde n}{\gamma \tilde p} \partial_{\tilde r}\tilde \nu
 \,\,\,\,, 
\eea
where
\bea\label{e-s}
\tilde E-\tilde S &=& 
\frac{\tilde G_{\rm eff}}{1+ 192\pi\xi^2\phi^2 \tilde G_{\rm eff}}
\left[ \frac{1}{A^2}(\partial_{\tilde r}\phi)^2 (1 + 12\xi) + 4\tilde 
V(\phi) + \tilde \rho -3 \tilde p 
+ 12\xi\phi{\partial \tilde V(\phi)\over \partial \phi} \right] \,\,\,\,,\\
\tilde E &=& 
\frac{\tilde G_{\rm eff}}{1+ 192\pi\xi^2\phi^2 \tilde G_{\rm eff}}
\left[-\frac{4\xi\phi 
(\partial_{\tilde r}\phi)(\partial_{\tilde r}\tilde \nu)}{A^2}
\left(1+ 192\pi\xi^2\phi^2 \tilde G_{\rm eff}\right) 
+\frac{1}{2A^2}(\partial_{\tilde r}\phi)^2 \left(1+8\xi+ 
64\pi\xi^2\phi^2 \tilde G_{\rm eff}\right) \right. \nonumber \\
\label{Eu}
& & \left. 
+ 4\xi\phi {\partial \tilde V(\phi)\over \partial \phi}+ 
\tilde V(\phi)\left(1 - 64\pi\xi^2\phi^2 \tilde G_{\rm eff}\right) 
+ \tilde \rho\left(1+128\pi\xi^2\phi^2 \tilde G_{\rm eff}\right) 
+ 192\pi\xi^2\phi^2 \tilde G_{\rm eff} \tilde p
\right] \,\,\,\,.
\eea
Here $\tilde E$ and $\tilde S$ are dimensionless as in 
Eq. (\ref{rotilde}), and we have dropped-out the tilde over $\phi$. 

We note that (\ref{matterfi}) with (\ref{lapsefi}) generalizes the 
Volkoff-Oppenheimer 
equation of hydrostatic equilibrium for the model (\ref{lag}). 
 We also note that the sources of the differential equations contain only
first order derivatives of the field variables and are thus suitable for
numerical integration with a Runge-Kutta algorithm. 

By subtracting Eq. (\ref{e-s}) from (\ref{Eu}) we can obtain an effective 
total pressure $\tilde S/3$ from which we recognize individual contributions, 
like the pressure terms associated with the scalar field 
\be
p_{\phi}= -\frac{1}{6A^2}(\partial_{\tilde r}\phi)^2 - \tilde V(\phi)\,\,\,,
\ee
and the NMC contributions: 
\bea
p^{\xi}_\phi &=& 
\frac{\tilde G_{\rm eff}}{3(1+ 192\pi\xi^2\phi^2 \tilde G_{\rm eff})}
\left[-\frac{4\xi\phi 
(\partial_{\tilde r}\phi)(\partial_{\tilde r}\tilde \nu)}{A^2}
\left(1+ 192\pi\xi^2\phi^2 \tilde G_{\rm eff}\right) 
\right. \nonumber \\
& & \left. 
+\frac{1}{2A^2}(\partial_{\tilde r}\phi)^2 \left(-1-16\xi+ 
64\pi\xi^2\phi^2 \tilde G_{\rm eff}\right) 
- 8\xi\phi {\partial \tilde V(\phi)\over \partial \phi}- 
\tilde V(\phi)\left(3 + 64\pi\xi^2\phi^2 \tilde G_{\rm eff}\right) 
\right] - p_{\phi} \,\,\,\,,\\
p^{\xi}_{\rm mat} &=&  
\frac{\tilde G_{\rm eff}}{3(1+ 192\pi\xi^2\phi^2 \tilde G_{\rm eff})}
\left[
128\pi\xi^2\phi^2 \tilde G_{\rm eff} \tilde \rho 
+3\tilde p(1+ 64\pi\xi^2\phi^2 \tilde G_{\rm eff} )
\right] -\tilde p \,\,\,\,\,.
\eea

For simplicity we will consider hereafter only the case $V(\phi)=0$ 
although the basic results are expected to remain qualitatively unchanged.

Regarding the equation of state, we make the standard assumption that the
nuclear matter in neutron stars is considered to be at zero temperature
(cold catalyzed matter) \cite{hww}. This matter is represented by an
equation of state usually parameterized by the baryon density number: 
\bea p= p(n)\,\,\,\,,\\ \rho= \rho(n) \,\,\,\,\,.  \eea In general the EOS
are not given analytically. In particular, for realistic EOS, i.e., those
built from effective field-theories or many-body nuclear calculations, the
above set of equations is given in tabulated form. Thus, a preliminary
interpolation is required. In this paper we will use the collection of EOS
of \cite{salgado1,salgado2} constructed with a logarithmic interpolation.
In particular we show results for three EOS representing the different
degrees of stiffness typically found in other models. These are the model
of Pandharipande in which only neutrons are taken into account \cite{pand}
(hereafter referred as to PandN; representative of a ``soft'' EOS), the
model II of D\'\i az-Alonso \cite{diaz} (hereafter referred as to DiazII; 
representative of a ``medium'' EOS), and the model ``0.17'' of Haensel et
al. \cite{hkp} (herafter referred as to HKP; representative of a ``stiff''
EOS). We point out that the DiazII and HKP models, being relativistic
models, are thus causal in all the domain of baryon densities. The PandN
model, despite being non-relativistic, is causal for the range of
densities usually found in stable configurations of neutron stars. We
refer the reader to the previous references, and the bibliography therein
for a more detailed description of such EOS.

\section{Boundary conditions and numerical methodology}
\bigskip

{\bf Interior solution}. 
The regularity condition at $r=0$ (center of the star) on the metric 
requires the boundary condition 
\be
m(0)= 0 \,\,\,.
\ee
The boundary condition on $\tilde \nu(r)$ is by definition 
[cf.Eq. (\ref{tildenu})]
\be
\tilde \nu (0) \equiv 0 \,\,\,\,.
\ee

The integration of Eq. (\ref{matterfi}) is performed by specifying 
the baryon density at the center of the star 
\be
n(0)= n_c \,\,\,\,,\\
\ee
with the regularity condition $\partial_r n(0)= 0$.

In a similar way the boundary and regularity conditions on the scalar
field are \bea \phi(0) &=& \phi_c \,\,\,\,,\\ \partial_r \phi (0) &=& 0
\,\,\,\,.  \eea On the other hand, the value $\phi_c$ cannot be arbitrary,
but must be so that $\phi$ satisfy the appropriate boundary conditions at
spatial infinite. This is enforced by the use of a standard shooting
method \cite{recipes}. 
 In the present paper, we only consider an asymptotically vanishing scalar
field. This corresponds to the situation in which spontaneous
scalarization might arise. The situation for a non-zero asymptotic value
of the scalar field can give rise to ``induced'' scalarization in the
neutron star, and can be also obtained in exactly the same fashion by a
shooting method. Figure {\ref{f:fi} shows the solution of the Klein-Gordon
equation satisfying vanishing asymptotic boundary conditions. Figure
\ref{f:firo} shows the ``trajectories'' $(\phi_c,\rho_c)$ for which those
conditions are verified within different neutron star configurations. 

{\bf Exterior solution.} 
We can compute the exterior solution by integrating the equations 
from the star surface to spatial infinite. 
Outside the support of the fluid variables $\rho, p$, we 
``compactify'' the space by a transformation on the $r$ coordinate of 
the kind  $u=1/r$, and 
integrate the resulting equations (see the Appendix) 
in the domain $u\in[1/R,0]$; at the star radius $R$  
(which is obtained numerically) and which is defined as the coordinate 
$r$ at which the pressure vanishes $p(R)=0 $, 
the values of the field variables are given numerically, and 
they represent the boundary conditions 
for the integration of the equations outside the star. The integration 
constants are fixed by matching continuously the interior with the 
(numerical calculated) exterior solution.

The physical lapse $N(r)= e^{\nu}$ at $r=0$ is calculated at the end of
the numerical integration from (\ref{tildenu})  \be \nu (0)= -\tilde
\nu_\infty \,\,\,\,, \ee where the value $\tilde \nu_\infty$ is obtained
from the numerical integration, and it ensures that at spatial infinite
our coordinates correspond to
 the standard Minkowski coordinates. Figure \ref{f:metric} depicts the metric 
potentials with asymptotic conditions matching with 
Minkowski coordinates at spatial infinite.
\bigskip

{\bf Global quantities.} The staticity of the configurations we are
studying ensures that the
 ADM (gravitational) mass and the Komar mass coincide, and can be easily 
evaluated from the integral 
\be
M_{\rm ADM}= {\rm lim}_{r\rightarrow\infty} \,\,
m (r) =  4\pi \int_{0}^{\infty} r^2 E(r) dr \,\,\,\,.
\ee

We emphasize that the energy density $E$ might not have compact support 
as it might include contributions of the scalar field. Therefore 
in order to compute the actual ADM mass   
the integration have to be performed from the center of the star 
to spatial infinite. 



Moreover, the conservation of the baryon number leads to the conserved
total baryon number of the star given by \be {\cal N} = 4\pi \int_{0}^{R}
n(r) A(r) r^2 dr \,\,\,\,\,.  \ee The total baryon mass is defined by \be
M_{\rm bar}= m_b {\cal N} \,\,\,\,\,, \ee where $m_b$ is the mean mass of
single baryons (1.66 $\times 10^{-27}$ kg).

As it is usual, we can define the total binding energy of the star as
\be
{\cal E}_{\rm bind}= M_{\rm ADM} - M_{\rm bar} \,\,\,\,,
\ee
and the fractional binding energy as
\be
{\cal E}_{\rm frac}= 1 - 
\frac{M_{\rm bar}}{M_{\rm ADM}} \,\,\,\,.
\ee

From the asymptotic behavior of the scalar field with vanishing boundary 
conditions,
\be
\phi(r) \sim \frac{G_0 \omega }{c^2 r} + {\cal O}(1/r^2)\,\,\,\,,
\ee
one can define the  so called ``scalar charge" $\omega$ as follows:
\be
\omega := - {\rm lim}_{\,\,\,r\rightarrow\infty} \left[r^2 \frac{c^2}{G_0}
\frac{d\phi}{dr}\right]\,\,\,\,.
\ee
Following \cite{damour1,damour2}, we introduce 
the coupling strength
\be\label{cs}
\alpha := -\frac{\omega}{M_{\rm ADM}}\,\,\,\,.
\ee

In the next section we provide the numerical analysis 
 of the above quantities, and their dependence
on the parameters $\rho_c= \rho(n_c)$, $\phi_c$, and the 
equation of state.

\bigskip
\section{Numerical results}

\subsection{Relativistic energetic analysis of spontaneous scalarization}
\bigskip In section III we showed that a naive Newtonian analysis leads to
the conclusion
 that the theory described by (\ref{lag}) with $\xi>0$ should not give
rise to a spontaneous scalarization since the presence of the scalar field
would tend to increase the total energy by decreasing of the negative
binding energy and thus to a configuration with a larger total energy than
the corresponding case in absence of scalar field, for a fixed baryonic
mass. In this section we will show that a numerical analysis of the full
relativistic energy contributions confirms that the phenomenon of
 spontaneous scalarization 
indeed occurs and that the naive analysis fails. 

Let us  separate the total energy density $E$ [see Eq. (\ref{Eu})] into
 contributions that can be identified as representing the different
effects that arise in the theory. This separation of course can be
consider to have no more than heuristic value, for even the energy density
itself has no invariant meaning in a diffeomorphism invariant theory. 
  We do this separation in the following way

\bea
\label{densint}
 \rho_{\rm int}(r) &:=& \rho - m_b n(r)   \,\,\,\,\,,\\
\rho^{\rm eff}_{\rm bar} (r) &:=& m_b n(r) A(r) \,\,\,\,\,\\
\rho^{\rm bind}_{\rm bar}(r) &:=& \rho^{\rm eff}_{\rm bar} (r)- m_b n(r) 
\,\,\,\,\,,\\
E_\phi &:=& \frac{1}{2A^2}(\partial_{ r}\phi)^2 \,\,\,\,,\\
E_\phi^\xi &:=& \frac{ G_{\rm eff}}{1+ 192\pi\xi^2\phi^2  G_{\rm eff}}
\left[-\frac{4\xi\phi 
(\partial_{ r}\phi)(\partial_{ r}\tilde \nu)}{A^2}
\left(1+ 192\pi\xi^2\phi^2  G_{\rm eff}\right) \right. \nonumber \\
&& \left. +\frac{1}{2A^2}(\partial_{ r}\phi)^2 \left(1+8\xi+ 
64\pi\xi^2\phi^2  G_{\rm eff}\right) \right] - E_\phi \,\,\,\,,\\
E_\rho^\xi &:=&
\frac{ G_{\rm eff}}{1+ 192\pi\xi^2\phi^2  G_{\rm eff}}
\left[ \rho\left(1+128\pi\xi^2\phi^2  G_{\rm eff}\right) 
+ 192\pi\xi^2\phi^2  G_{\rm eff}  p
\right] -  \rho\,\,\,\,.
\label{densroxi}
\eea
Then it is clear that 
\be
E=   \rho^{\rm eff}_{\rm bar} -  
\rho^{\rm bind}_{\rm bar} +  \rho_{\rm int} + 
     E_\phi  +  E_\phi^\xi + E_\rho^\xi \,\,\,\,.
\ee
This decomposition has the following interpretation:
$\rho^{\rm eff}_{\rm bar}$ is the physical baryon density, that is, 
the energy density that will give rise to the total baryon mass;
 $\rho_{\rm int}$ is the interaction energy between baryons (usually it is
given in a complicated way related to the physics model of the nuclear
matter), $- \rho^{\rm bind}_{\rm bar}$ is the negative standard binding
energy-density of baryons (is the part that is associated only with the
difference of the baryon density multiplied by the geometrical factor
corresponding to the proper volume element on spacelike hypersurfaces
$\Sigma_t$ and the baryon density itself. The integral of this difference
is what yields the total binding baryon energy in the case of pure general
relativity \cite{wald}), $E_\phi$ is the energy-density of the scalar
field as it is usually defined without NMC, $E_\phi^\xi$ is the
contribution to the energy-density of the scalar field due to the
non-minimimal coupling (note that this vanishes when $\xi=0$), and
finally, $E_\rho^\xi$ is the contribution to the total matter-energy
density due to the presence of the non-minimal coupling (we also note that
it vanishes for $\xi=0$). The integration of each of the terms from $r=0$
to $r= +\infty$ will give the total energy contributions:  
\be\label{madm}
M_{\rm ADM}= M_{\rm bar} + M_{\rm int} - M_{\rm bind} + M_\phi +
                M_\phi^\xi + M_\rho^\xi \,\,\,\, .
\ee
We remark that the above interpretation is sustainable as far as 
we concern ourselves with the functional form of the terms themselves 
and regard them as applicable to different test configurations. On the 
other hand the particular configuration that minimizes the total 
$M_{\rm ADM}$ arises from an interplay of all such effects and therefore 
the actual value of each of the terms in Eq. (\ref{madm}) is 
affected by the presence of the other terms. Nevertheless the 
heuristic value of the arguments is validated since such affectations 
can be consider as higher order effects.

In the absence of scalar field we have
\be
M_{\rm ADM}^{\rm GR}= M_{\rm bar}^{\rm GR} + M_{\rm int}^{\rm GR} - 
M^{\rm GR}_{\rm bind}\,\,\,\,.
\ee
So if we want to compare two configurations at fixed baryon mass 
($M_{\rm bar}= M_{\rm bar}^{\rm GR}$),  
one with $\phi\neq 0$ and the other with null scalar field, we have
\bea
\Delta M &:=& M_{\rm ADM} - M_{\rm ADM}^{\rm GR} \nonumber \\
       &=& M_{\rm int}- M_{\rm int}^{\rm GR} - 
         \left(M_{\rm bind}- M^{\rm GR}_{\rm bind}\right)
            + M_\phi + M_\phi^\xi + M_\rho^\xi \,\,\,\,.  \eea Now
according to the Newtonian intuition we expect that $|M^{\rm GR}_{\rm
bind}| > |M_{\rm bind}| $. So the term $-\left(M_{\rm bind}- M^{\rm
GR}_{\rm bind}\right)$ would be positive. Moreover the energy $M_\phi$ is
also positive and we expect that $M_{\rm int}- M_{\rm int}^{\rm GR}$ is
small. In order to have spontaneous scalarization we need $\Delta M <0$
and so the term $M_\phi^\xi + M_\rho^\xi$ should be negative enough to
more than compensate for the positive energy contributions and also for
the decrease in the ordinary binding energy.  Figure \ref{f:nrjs} shows
the local behavior of such energy contributions (referred with lower-case
letters in the caption) for a configuration with spontaneous
scalarization, and confirms the previous conjectures. Figure
\ref{f:nrjsden} depicts the corresponding energy-densities as defined by
the Eqs. (\ref{densint})-(\ref{densroxi}). Table I shows examples of
configurations exhibiting spontaneous scalarization as compared with those
in absence of scalar field at the same baryon mass. We can appreciate from
table I that the different energy contributions add to reproduce the ADM
mass and that the value for the configuration with $\phi\neq0$ is lower
than its corresponding value with $\phi=0$. Note also that the binding
baryon energies in absence of scalar field are smaller than their
corresponding in the configurations with $\phi\neq0$. This confirms the
original Newtonian intuition. However, the relativistic energetic analysis
also shows that the Newtonian naive argument failed to explain the
spontaneous scalarization with $\xi>0$ since it did not take into account
the large negative energy contribution $M_{\rho}^\xi$ which is the
responsible for the decrease of the total energy ($M_\phi^\xi$ turns to be
always positive), and as it is shown in the examples of table I, is the
dominant part of the above difference $\Delta M$. This contribution as
well as $M_{\phi}^\xi$ have no Newtonian counterpart.

\subsection{Effects of the equations of state}
\bigskip
   
A simple and enlightening way to present the effects of the equation of
state on the phenomena 
behavior of the other global quantities as a function of the total baryon
mass of the neutron star. Figures \ref{f:alpha} and \ref{f:alpha2} show
the behavior of the coupling strength $\alpha$ [see Eq. (\ref{cs}] as a
function of the total baryon mass for three EOS. We note that there are
critical values $M^{\rm crit}_{\rm bar}$ beyond which the phenomenon of
spontaneous scalarization ensues. The critical values correspond to the
point at which configurations with a non-vanishing scalar field become
energetically more favorable than the configurations with $\phi(r)=0$.
This behavior can be better appreciated in Fig. \ref{f:bind}. It is
interesting to note that $M^{\rm crit}_{\rm bar}$ is an increasing
function of the stiffness of the EOS (for example when $\xi=2$, $M^{\rm
crit}_{\rm bar}$ for the HKP EOS increases in $\sim 37$\% relative to that
for the PandN EOS.), but that the maximum values of $\alpha$ seem to be
almost independent of the EOS.

Figure \ref{f:alpha2} shows that the larger the value of $\xi$ the less
important the effect of the EOS on the critical value $M^{\rm crit}_{\rm
bar}$. This can be understood on basic grounds by looking at the
expression of $E_{\rho}^\xi$ [Eq. (\ref{densroxi})], which as we discussed
is the main responsible for spontaneous scalarization. As $\xi$ increases
$E_{\phi}^\xi \rightarrow -\rho$. This means that its relative
contribution is becoming dominant regardless of the central-value of
$\rho$. Therefore spontaneous scalarization can take place for smaller
values of $\rho_c$. Furthermore, the lower the value $\rho_c$ the more
alike are the equations of state since the three EOS are built with the
same standard neutron-star-crust EOS which corresponds to that of
\cite{eos}. So the configurations of the star with lower central values of
$\rho$ are very much alike. This is the reason for the critical baryon
masses look very similar for different EOS with large values of $\xi$.
Table \ref{t:crit} shows the values of quantities corresponding to the
onset of spontaneous scalarization (i.e., corresponding to the critical
$M_{\rm bar}$).

As it was pointed out in \cite{damour2}, the non-perturbative effects of the 
scalar field in the neutron star are the analogue of spontaneous magnetization 
in ferromagnets at temperatures lower than the Curie point. In the Landau 
ansatz where the free energy $\Gamma$ of the ferromagnet is a function of the 
magnetization $\mu$ in the form $\Gamma= a(T-T_C) \mu^2/2 + b\mu^4/4$, 
it appears two non-trivial minima for temperatures $T<T_C$. The analogue to 
the curve giving the locus of the minima of $\Gamma$, is depicted for the case 
of spontaneous scalarization in figure \ref{f:ss}, where the scalar charge 
$\omega$ plays the role of $\mu$, the total mass $M_{\rm ADM}$ the role 
of $\Gamma$, and finally the baryon mass is the analogue of the 
temperature. Obviously in this case, the complexity of the system makes 
difficult an attempt to provide an analytic expression of the form 
$M_{\rm ADM}= M_{\rm ADM} (\omega,M_{\rm bar})$ for which we could recover 
the behavior of figure \ref{f:ss}.

Another effect related to the phenomenon of spontaneous scalarization is
the increase of the maximum masses (both $M_{\rm ADM}$ and $M_{\rm bar}$)
as compared with the values allowed by general relativity.  Figure
\ref{f:mvsrho} shows the dependence of $M_{\rm ADM}$ on $\rho_c$ for the
selected EOS. The behavior of these profiles are qualitatively the same as
in general relativity (see Ref. \cite{salgado1}). Table \ref{t:max} gives
the relevant properties of the neutron star models corresponding to the
maximum mass configurations. Note the increase of $M_{\rm ADM}$ with
$\xi$. 

\bigskip
\section{Discussion}

One of the main goals of this work was to understand the discrepancy
between the Newtonian expectations and the relativistic energetic
analysis. A {\it fortiori} it turns out that the difference can be
understood by noting the different roles that the gravitational constant
plays in the energy considerations in the Newtonian
 and the relativistic theories. As discussed in section III, the Newtonian
point of view leads to the expectation that the total mass should be
expressed in the form:  \be {\cal M} = \int \rho_{\rm bar} (\overline x)
d^3 \overline x + \int \rho_\phi (\overline x) d^3\overline x - G_{\rm
eff} (\phi) \int \frac {\rho_{\rm bar} (\overline x) \rho_{\rm bar}
(\overline y)}{|\overline x- \overline y|} d^3\overline x d^3\overline y
+... \,\,\,\,, \label{Newt} \ee so that the gravitational constant appears
as the coefficient of a term that represents the binding energy which is
negative, so a decrease in $G_{\rm eff}$ results
 in an increase on ${\cal M}$.
 In the relativistic theory the ADM mass itself is of gravitational origin
because it is a measure of the departure of the metric (at large
distances)  from the Minkowski metric, i.e., it represents the $1/r$
coefficient of such deviation. In this sense we can say that the only
thing which has meaning is the combination $G\times {\cal M}$ and not
${\cal M}$ itself. Thus the quantity that must be minimized is
(heuristically):  
\be G{\cal M} = G_{\rm eff}(\phi) \left[
 \int  \rho_{\rm bar} (\overline x) d^3 
\overline x  + \int  \rho_\phi (\overline x) d^3\overline x - 
G_{\rm eff} (\phi) \int   
\frac {\rho_{\rm bar} (\overline x) \rho_{\rm bar} (\overline y)}{|\overline x- 
\overline y|} d^3\overline x d^3\overline y +... \,\,\,\,,
\right]
\label{(Rel1)}
\ee
 Which we can write as
\be
G {\cal M} = G_{\rm eff}(\phi) {\cal M}_{\rm bar}  + 
G_{\rm eff}(\phi) {\cal M}_{\phi} 
-G_{\rm eff}^2 (\phi)  \int   
\frac {\rho_{\rm bar} (\overline x) \rho_{\rm bar} (\overline y)}{|\overline x- 
\overline y|} d^3\overline x d^3\overline y +... \,\,\,\,,
\label{(Rel2)}
\ee

The point is that the appearance of a nonzero value of $\phi$ which
results in the second term becoming positive (from a zero value when
$\phi=0$) can be more than compensated by the decrease in the first term
due to the decrease in $G_{\rm eff}$ that is associated with an increase
in $\phi$. The change in the third term, which in effect is in the
direction inferred from the Newtonian analysis, playing now a secondary
role. This explains why the Newtonian analysis failed and why the
relativistic analysis shows that the phenomenon of spontaneous
scalarization is associated with an effective gravitational constant that
decreases with $\phi$. This explanation is supported by the numerical
analysis since the term $M_{\rho}^\xi$ which represents precisely the
effect described here turns out to be the one that is responsible for the
lowering of the ADM mass of the configuration with a scalar field in
comparison to the configuration without one. 

 The second goal of this paper was to study the dependence of the
parameters associated with the spontaneous scalarization with the equation
of state of the nuclear matter. This can be summarized as follows. First
the critical baryon masses at which spontaneous scalarization develops
increases as the equation of state becomes more rigid. As $\xi$ increases
those critical values become much less dependent of the EOS. Second, the
maximum values of the coupling strength $\alpha$ seem to be independent of
the EOS. Finally, the effect of the scalar field makes the neutron star to
support larger masses than in general relativity, and this is more evident
as $\xi$ increases. 



\medskip
\section{Acknowledgments}
\medskip

We wish to thank H. Quevedo
for helpful discussions.
M.S and D.S. would like to acknowledge partial support
from DGAPA-UNAM project IN105496 and to thank the supercomputing department of 
DGSCA-UNAM.

\newpage
\begin{appendix}
\section*{Dimensionless form of field equations at the exterior}
\bigskip
By performing a ``conformal'' transformation of the kind $u=1/\tilde r$ 
in Eqs. (\ref{massu})--(\ref{Eu}) (with $\tilde \rho= \tilde p = 
V(\phi)= 0$) we obtain the equations to be solved 
in the domain $u\in[1/R,0]$ which corresponds to the range from the 
radius of the star to spatial infinite.

\bea
\partial_u \tilde m &=&  -4\pi \tilde E_u \,\,\,\,\,,\\
\label{lapsefiu}
\partial_u\tilde \nu  &=& - \frac{(1-2\tilde m u)^{-1}}{1-16\pi\xi 
u \phi (\partial_u\phi) \tilde G_{\rm eff} }
 \left\{\tilde m  + 4\pi \tilde G_{\rm eff}(1-2\tilde m u)\left[
\frac{u}{2}(\partial_u\phi)^2 + 8\xi\phi\partial_u\phi
\right]  \right\} \,\,\,\,, \\
\partial_{uu} \phi &=& 
-\left[ \partial_u \tilde \nu- 
\left(1- 2\tilde m u\right)^{-1}
\left(\tilde m - 4\pi u \tilde E_u\right) 
\right] \partial_u \phi \nonumber \\
& & 
-\frac{16\pi\xi\phi \tilde G_{\rm eff}}{1+ 192\pi\xi^2\phi^2 \tilde G_{\rm eff}}
 (\partial_u \phi)^2 (1 + 12\xi)  \,\,\,\,,\\
\tilde E_u &=& 
\frac{\tilde G_{\rm eff} (1-2\tilde m u)}
{1+ 192\pi\xi^2\phi^2 \tilde G_{\rm eff}}
\left[- 4\xi\phi 
(\partial_u\phi)(\partial_u\tilde \nu)
\left(1+ 192\pi\xi^2\phi^2 \tilde G_{\rm eff}\right) \right.\nonumber \\
& & \left. + \frac{1}{2}(\partial_u\phi)^2 \left(1+8\xi+ 
64\pi\xi^2\phi^2 \tilde G_{\rm eff}\right) \right] \,\,\,\,.
\eea
\end{appendix}

\begin{table*}
\caption[]{\label{t:spon} Examples of configurations exhibiting 
spontaneous scalarization compared with those at 
the same baryon mass in absence of scalar field. 
$\xi$ is the non-minimal coupling constant; $M_{\rm bar}$ is the total 
baryon mass; $M_{\rm bin}$ is the baryon-binding energy; 
$M_{\rm int}$ is the total baryon-interaction energy;  
$M_\phi$ is the total mass contribution due to the single scalar field; 
$M_\rho^\xi$ is the mass contribution of the   
correction to the total matter-energy by the presence of the 
non-minimal coupling; $M_\phi^\xi$ is the mass contribution of the   
correction to $M_\phi$ by the presence of the non-minimal coupling; 
$M_{\rm ADM}$ is the ADM-mass. 
We use $M_\odot= 1.989 \times 10^{30}\ {\rm kg}$. }
\begin{flushleft}
\begin{tabular}{c|ccccccccc}
\hline\noalign{\smallskip} 
$\displaystyle {{\rm EOS} \atop \ }$ &
$\displaystyle {\xi \atop \ }$ &
$\displaystyle {\phi(r) \atop \ }$ &
$\displaystyle {M_{\rm bar} \atop [M_\odot]}$ &
$\displaystyle {M_{\rm bind} \atop [M_\odot]}$ &
$\displaystyle {M_{\rm int} \atop [M_\odot]}$ &
$\displaystyle {M_{\phi} \atop [M_\odot]}$ &
$\displaystyle {M_{\rho}^{\xi} \atop [M_\odot]}$ &
$\displaystyle {M_{\phi}^{\xi} \atop [M_\odot]}$ &
$\displaystyle {M_{\rm ADM} \atop [M_\odot]}$ \\
\noalign{\smallskip}
\hline\noalign{\smallskip}
DiazII   & 2 & $\neq 0$ & 1.563 & -0.179 &0.082 &0.011 & -0.226 & 0.179 
         & 1.429 \\
         & 2 & $= 0$ & 1.563 & -0.207 &0.077 &0 & 0 & 0 & 1.433 \\
          & 6 & $\neq 0$ & 2.170 & -0.320 &0.170 &0.011 & -0.559 & 0.413 
          & 1.885 \\
          & 6 & $= 0$ & 2.170 & -0.469 &0.204 &0 & 0 & 0 & 1.904 \\
\noalign{\smallskip}
\hline
\end{tabular}
\end{flushleft}
\end{table*}

\begin{table*}
\caption[]{\label{t:crit} Critical configurations towards spontaneous 
scalarization. $n_{\rm c}$, $\rho_{\rm c}$ are the central baryon and proper 
energy densities respectively, $p_{\rm c}$ is the central pressure, 
$R$ is the 
radius of the star, $N_{\rm c}$ is the value of the lapse at the center of 
the star. The remaining symbols are defined in the caption of 
Table \ref{t:spon}. We use
$G_0=6.672\times 10^{-11}\ {\rm m}^3{\rm kg}^{-1}{\rm s}^{-2}$,
$c=2.9979\times 10^8\ {\rm m\, s}^{-1}$,
$M_\odot= 1.989 \times 10^{30}\ {\rm kg}$, and $\rho_{\rm nuc}= 1.66\times
 10^{17} {\rm kg\, m^{-3}}$. }
\begin{flushleft}
\begin{tabular}{c|cccccccccc}
\hline\noalign{\smallskip}
 $\displaystyle {{\rm EOS} \atop \ }$  &
 $\displaystyle { \xi \atop \ }$  &
 $\displaystyle {n_{\rm c} \atop [.1{\rm fm}^{-3}] }$ &
 $\displaystyle {\rho_{\rm c}\atop [\rho_{\rm nuc} c^2]}$  &
 $\displaystyle {p_{\rm c}\atop [\rho_{\rm nuc} c^2]}$  &
 $\displaystyle {M_{\rm ADM} \atop [M_\odot]}$ &
 $\displaystyle {M_{\rm bar}\atop [M_\odot]}$  &
 $\displaystyle {R \atop [{\rm km}]}$  &
 $\displaystyle {1- \frac{M_{\rm bar}}{M_{\rm ADM}} \atop \ }$  &
 $\displaystyle {\frac{2G_0M_{\rm ADM} }{c^2 R} \atop \ }$  &
 $\displaystyle {N_c \atop \ }$  \\
\noalign{\smallskip}
\hline\noalign{\smallskip}
PandN  & 2 &4.22 & 4.46 & 0.254 & 0.596 &0.617 & 11.29 
         &-0.035 & 0.156 & 0.822 \\
         & 6 &1.76 &1.80 &0.026 &0.1779 & 0.1779 &18.65 & -5.87
$\times 10^{-5}$ &0.028 &0.944   \\ \hline
DiazII   & 2 & 2.52 &2.65 &0.150 & 0.757 &0.787 &13.90  
         &-0.038 &0.160 &  0.824\\
         &6 & 1.17 & 1.20  & 0.018 & 0.2077 &0.2079 & 20.08 
         &-0.0012 & 0.030  &  0.945   \\ \hline
HKP      & 2 &1.97 &2.05 &0.122 & 0.811 &0.845 &14.24 
         &-0.042 &0.169 & 0.826 \\
      & 6 &1.243 &1.27 &0.020 &0.1970 & 0.1973 & 18.98 &-0.0012 &
       0.030 & 0.945  \\    
\noalign{\smallskip}
\hline
\end{tabular}
\end{flushleft}
\end{table*}

\begin{table*}
\caption[]{\label{t:max} Maximum mass models.
 $\phi_{\rm c}$ is the central 
scalar field,
$M_{\rm s}$ is the ``gravitational mass'' at the surface of 
the star, 
 $\omega$ is the scalar charge, $\alpha$ is the 
coupling strength between the scalar field and gravitation. 
The remaining identifiers are defined in the caption of Tables \ref{t:spon} 
and \ref{t:crit}. The configurations in absence of scalar field corresponds 
to the maximum mass models in general relativity.}
\begin{flushleft}
\begin{tabular}{c|cccccccccccccc}
\hline\noalign{\smallskip}
 $\displaystyle {{\rm EOS} \atop \ }$  &
 $\displaystyle { \xi \atop \ }$  &
 $\displaystyle {n_{\rm c} \atop [.1{\rm fm}^{-3}] }$ &
 $\displaystyle {\rho_{\rm c}\atop [\rho_{\rm nuc} c^2]}$  &
 $\displaystyle {p_{\rm c}\atop [\rho_{\rm nuc} c^2]}$  &
 $\displaystyle {\phi_{\rm c}\atop }$  &
 $\displaystyle {M_{\rm s} \atop [M_\odot]}$ &
 $\displaystyle {M_{\rm ADM} \atop [M_\odot]}$ &
 $\displaystyle {M_{\rm bar}\atop [M_\odot]}$  &
 $\displaystyle {R \atop [{\rm km}]}$  &
 $\displaystyle {\omega \atop [M_\odot]}$ &
 $\displaystyle {1- \frac{M_{\rm bar}}{M_{\rm ADM}} \atop \ }$  &
 $\displaystyle {\frac{2G_0M_{\rm ADM} }{c^2 R} \atop \ }$  &
 $\displaystyle {\alpha \atop \ }$ &
 $\displaystyle {N_c \atop \ }$  \\
\noalign{\smallskip}
\hline\noalign{\smallskip}
PandN         & 0 &17.36 &24.69 &11.69 &0 &1.662 &1.662 & 1.932&8.52 &0
        &-0.162 &0.575 &0 &0.310 \\
       & 2 &15.51 &20.98 & 8.73 & 0.0171 &1.620 & 1.694 &1.981 &8.91 
       &0.0536 &-0.169 & 0.558 & -0.0316 & 0.344 \\
         & 6 &15.51 &20.98 &8.73 &0.0169 &1.570 &1.776 &2.118 &9.13 &0.0830
         &-0.192 &0.574 &-0.0467 &0.337 \\ \hline
DiazII   & 0 & 11.14 &15.20 &5.30 & 0 &1.933 &1.933 &2.210 &10.92 &0 
         &-0.143 &0.522 &0 & 0.375 \\
         &2 & 10.75 & 14.49 & 4.90 & 0.0226 & 1.868 & 1.990 & 2.297 
          & 11.21&  0.0818 & -0.153 & 0.524 &-0.0411 & 0.380 \\
          & 6 &10.94 &14.85 &5.10 &0.019 &1.801 &2.085 &2.451 &11.42
          &0.115 &-0.175 & 0.538 &-0.055 &0.368 \\ \hline
HKP       & 0 & 6.32 &8.76 &4.29 &0 &2.835 &2.835 &3.421 &13.67 &0
          &-0.206 &0.612 &0 &0.301 \\
    &2 &5.64 &7.42 &3.27 & 0.0146 & 2.768 & 2.879 &3.492 & 14.15
      &0.0805 &-0.212 &0.600 & -0.0279 & 0.332 \\
  & 6 &5.56 &7.28 &3.16 &0.015 &2.667 &3.018 &3.732 &14.56 
     &0.132 &-0.236 & 0.612 & -0.043& 0.329 \\
\noalign{\smallskip}
\hline
\end{tabular}
\end{flushleft}
\end{table*}

\begin{figure*} 
\vspace{1cm} 
\psfig{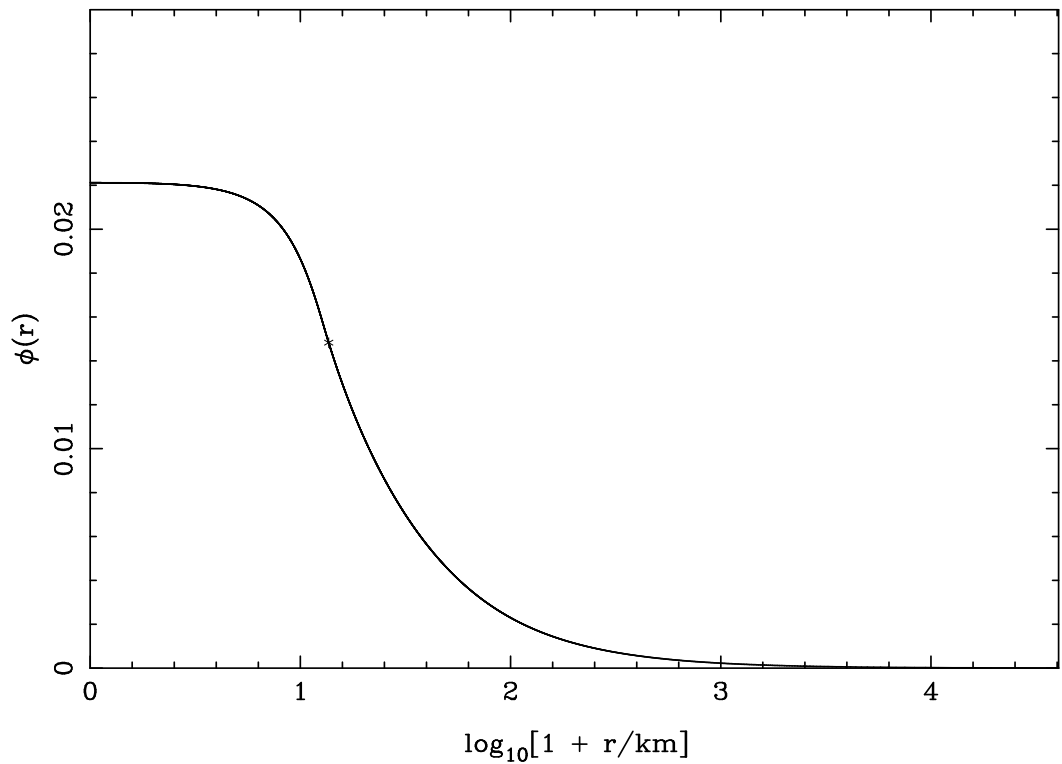} 
\hspace*{-2.5in} 
\caption[]{\label{f:fi} 
Scalar-field as a function of $r$ for $\xi=6$,
$n_{\rm c} \sim 0.7\ {\rm fm}^{-3}$ ($\rho_{\rm c} \sim 8.37\ \rho_{\rm
nuc} c^2$)  and $\phi_{\rm c}\sim 0.0221$ for the EOS DiazII. The asterisk
depicts the location of the star surface ($R\sim 12.63$ km). The
computations includes the domain from $r=0$ to $r=+\infty$, however for
pictorial convenience here the logarithmic scale shows the plot only to
some extent which includes various orders of magnitude beyond the star
surface.} 
\end{figure*} 
\vskip 1cm

\begin{figure*}
\vspace{1cm}
\psfig{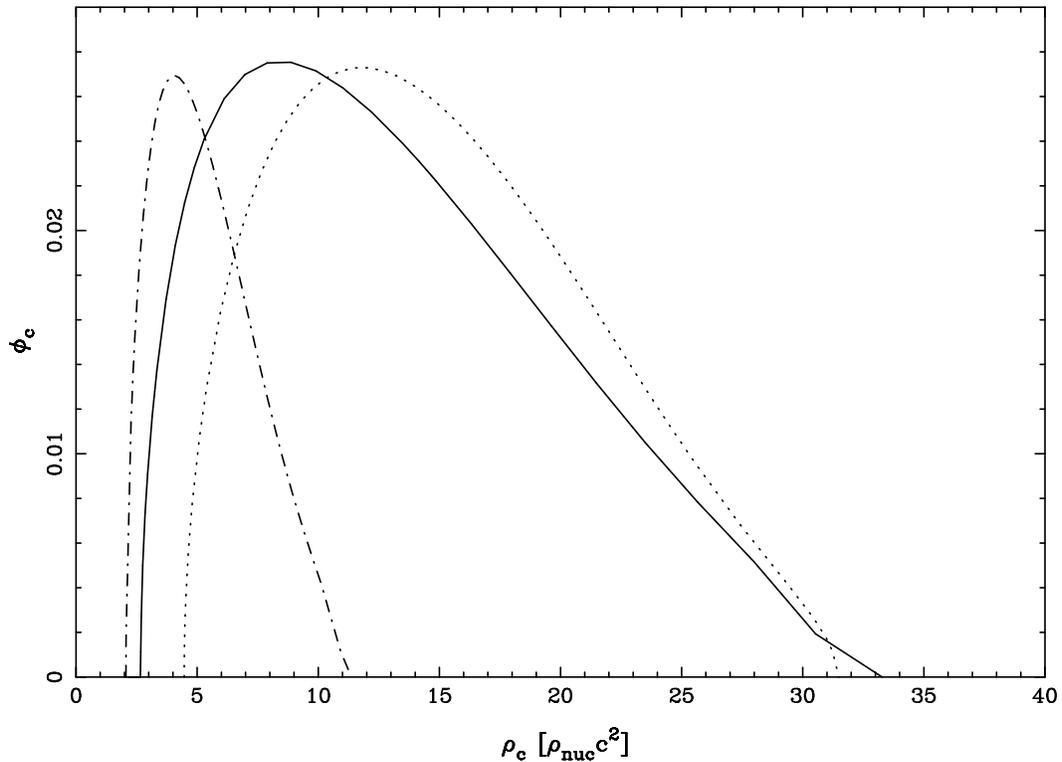}
\hspace*{-2.5in}
\caption[]{\label{f:firo}
Functional dependence of $\phi_c$ vs $\rho_c$ obtained from a 
shooting method with $\xi=2$ for three different equations of state: 
HKP (dash-doted line), DiazII (solid line) and PandN (dotted line). 
The appearence of the non-trivial solutions ($\phi_c\neq 0$) 
marks the onset of spontaneous scalarization.}
\end{figure*}
\vskip 1cm

\begin{figure*} \vspace{1cm} \psfig{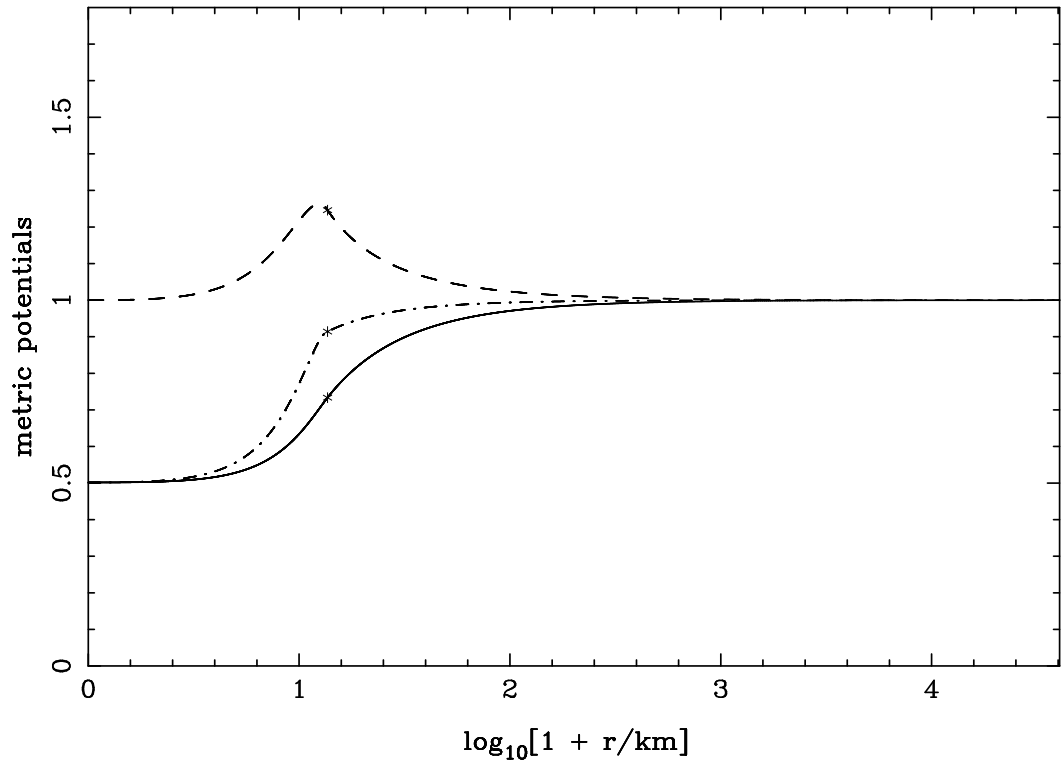} \hspace*{-2.5in}
\caption[]{\label{f:metric} Metric potentials as a function of $r$
corresponding to the case of Fig. \ref{f:fi}. The curves refer to the
lapse function $N(r)$ (solid line), to the metric potential $A(r)$ defined
by Eq. (\ref{AA}) (dashed line), and to the product $AN$ (dash-dotted
line). Note that unlike the Schwarzschild exterior solution this product
differs from unit outside the star. Here the asterisks indicates the
location of the star surface.} 
\end{figure*} \vskip 1cm

\begin{figure*}
\vspace{1cm}
\psfig{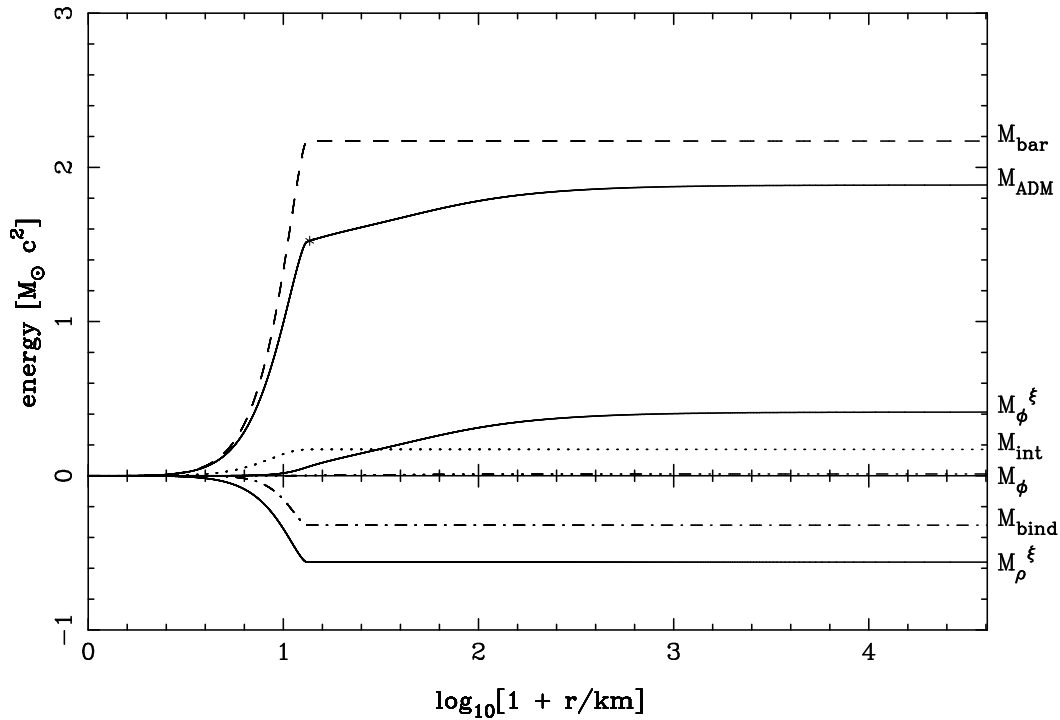}
\hspace*{-2.5in}
\caption[]{\label{f:nrjs}
Energies for the same case of Fig. \ref{f:fi} as a function 
of $r$. The profiles represent the 
integration of the energy-densities of Fig. \ref{f:nrjsden} 
from $r=0$ to $\infty$. 
The upper solid lines (positive values) correspond to $m(r)$ 
 (larger values) and $m_\phi^\xi (r)$ (smaller values) respectively; 
the dashed line stands for $m_{\rm bar}(r)$; the dotted line corresponds to 
$m_{\rm int}(r)$; the dash-dotted line (positive values) corresponds to 
$m_\phi(r)$; the dash-dotted line (negative values) stands for 
$m_{\rm bind}(r)$; the lower solid line (negative values) 
refers to $m_\rho^\xi(r)$. The asymptotic values of these profiles 
correspond to those of table \ref{t:spon}. Note that 
the mass $m(r)$ continues to grow outside the star 
(the asterisk depicts the star surface) due to the contribution of 
the energies associated with the scalar field.}
\end{figure*}
\vskip 1cm

\begin{figure*}
\vspace{1cm}
\psfig{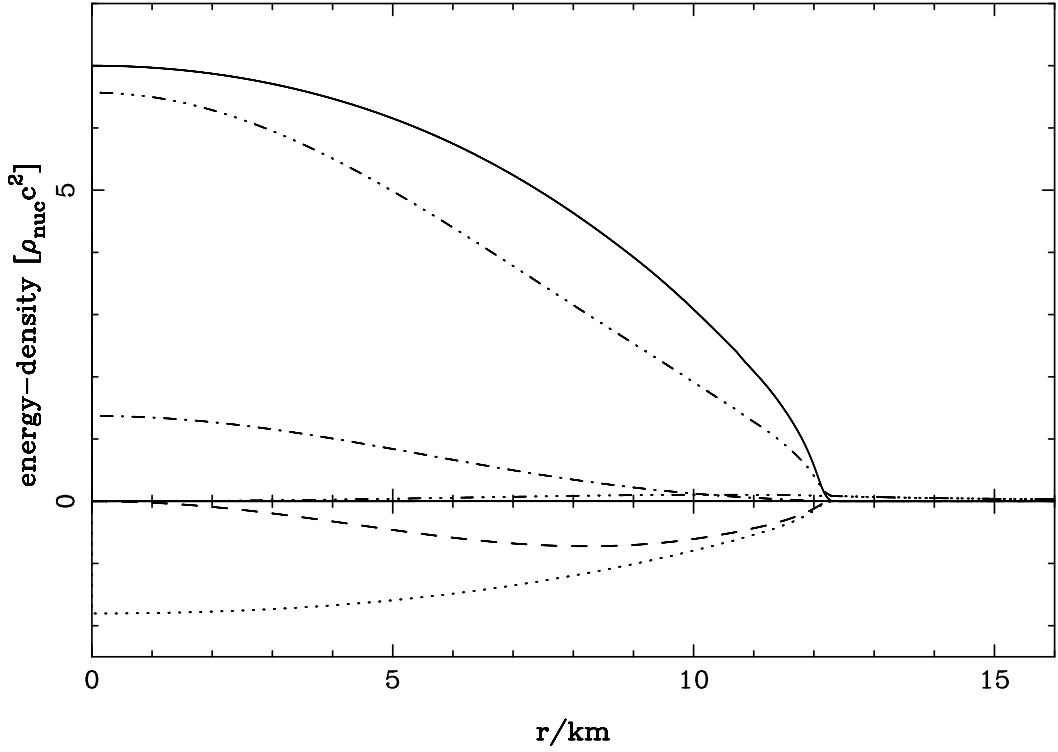}
\hspace*{-2.5in}
\caption[]{\label{f:nrjsden}
Energy-densities as a function of $r$ for the same case of Fig. \ref{f:fi} 
(see the text for description) corresponding to (from the upper to the 
lower curve respectively) 
$\rho^{\rm eff}_{\rm bar}$ (solid line), $E$ (dash-dotted line; larger values), 
$\rho_{\rm int}$ (dash-dotted line; lower values), 
$E^\xi_\phi$ (dash-dotted line; lowest positive values), 
$\rho^{\rm bind}_{\rm bar}$ (dashed line), and  $E^\xi_\rho$ (dotted line). 
In this scale, the energy-density $E_\phi$ lies almost on the r-axis.}
\end{figure*}
\vskip 1cm

\begin{figure*} \vspace{1cm} \psfig{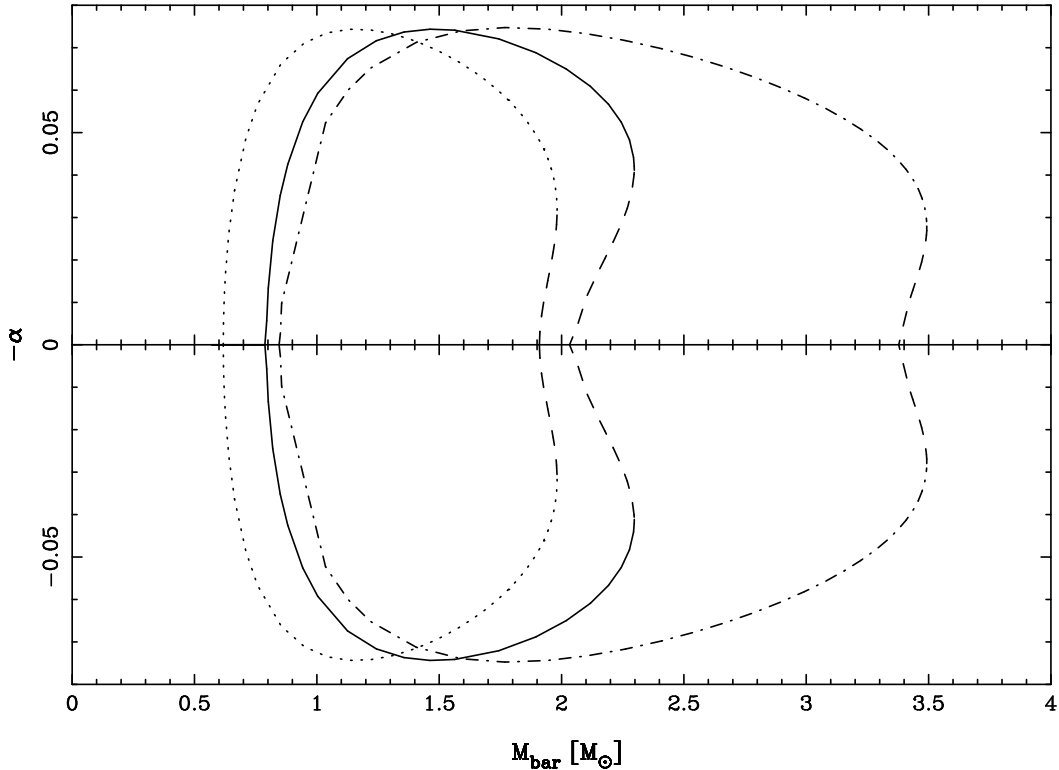} \hspace*{-2.5in}
\caption[]{\label{f:alpha} Coupling strength vs the total baryon mass for
the EOS labeled HKP (dash-dotted line), DiazII (solid line), and PandN
(dotted line) for $\xi=2$. The dashed segments of the lines refer to the
unstable configurations. The positive branches ($-\alpha>0$)  correspond
to configurations with $\phi>0$, while the negative ones ($-\alpha<0$)
refer to those with $\phi<0$. The critical baryon masses for spontaneous
scalarization are shown in Table \ref{t:crit}. } \end{figure*} \vskip 1cm

\begin{figure*}
\vspace{1cm}
\psfig{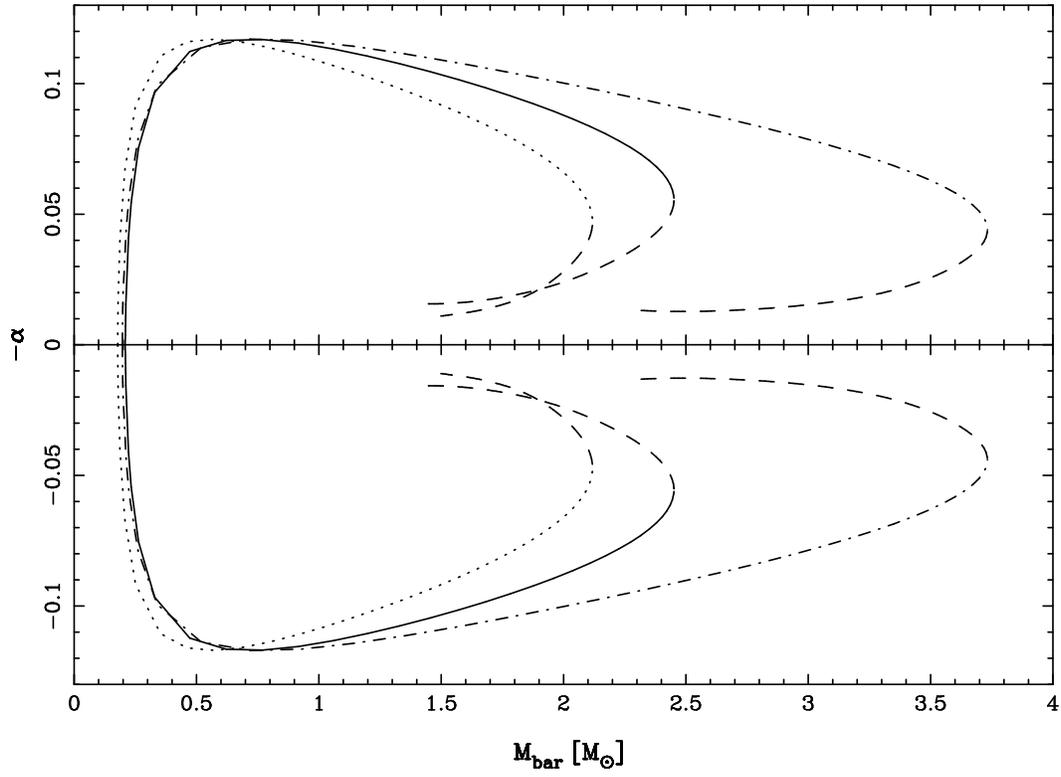}
\hspace*{-2.5in}
\caption[]{\label{f:alpha2}
Similar as Fig. \ref{f:alpha} for $\xi=6$. }
\end{figure*}
\vskip 1cm

\begin{figure*}
\vspace{1cm}
\psfig{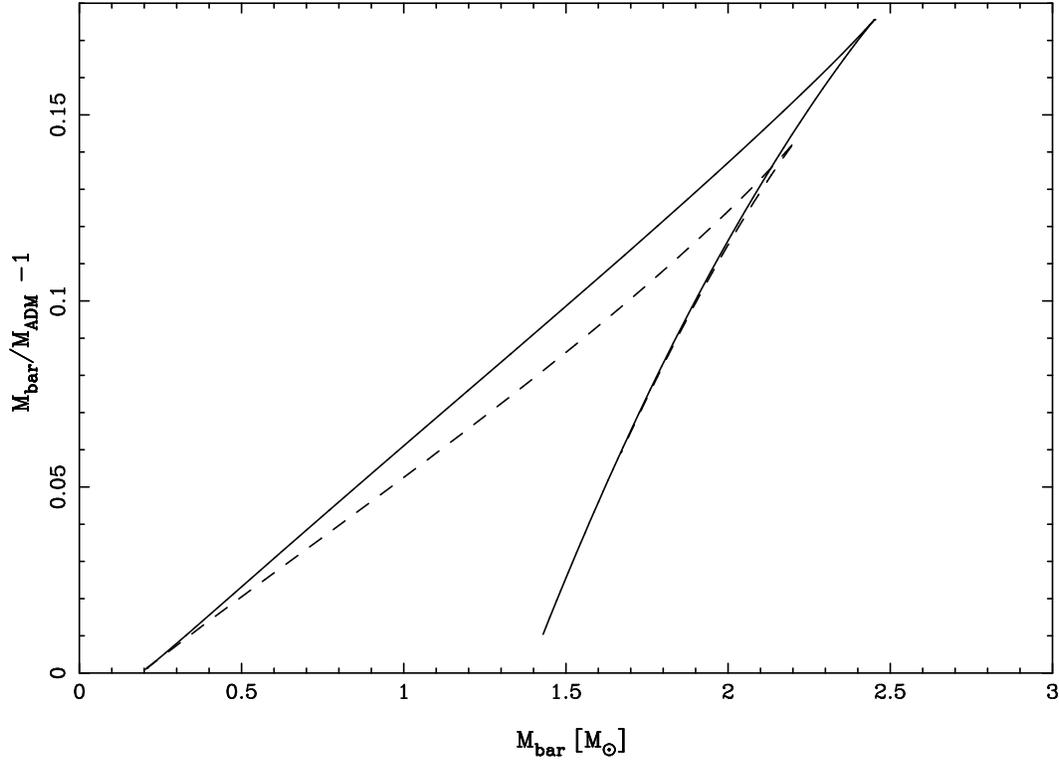}
\hspace*{-2.5in}
\caption[]{\label{f:bind}
Total fractional binding-energy as a function of the total baryon mass for 
the EOS DiazII with $\xi=6$. The solid line refers to 
configurations computed with $\phi\neq 0$  while the 
dashed line depicts the configurations with $\phi=0$. 
Here second branches (unstable configurations) are also shown. 
Beyond some critical baryon-mass the configurations with $\phi\neq 0$ are 
energetically more favorable than the corresponding at the same baryon mass 
with $\phi=0$. Note the increase in the maximum baryon-mass from 
$2.21\,M_\odot$ ($\phi= 0$) to $2.45\,M_\odot$ ($\phi\neq 0$).}
\end{figure*}
\vskip 1cm

\begin{figure*} \vspace{1cm} \psfig{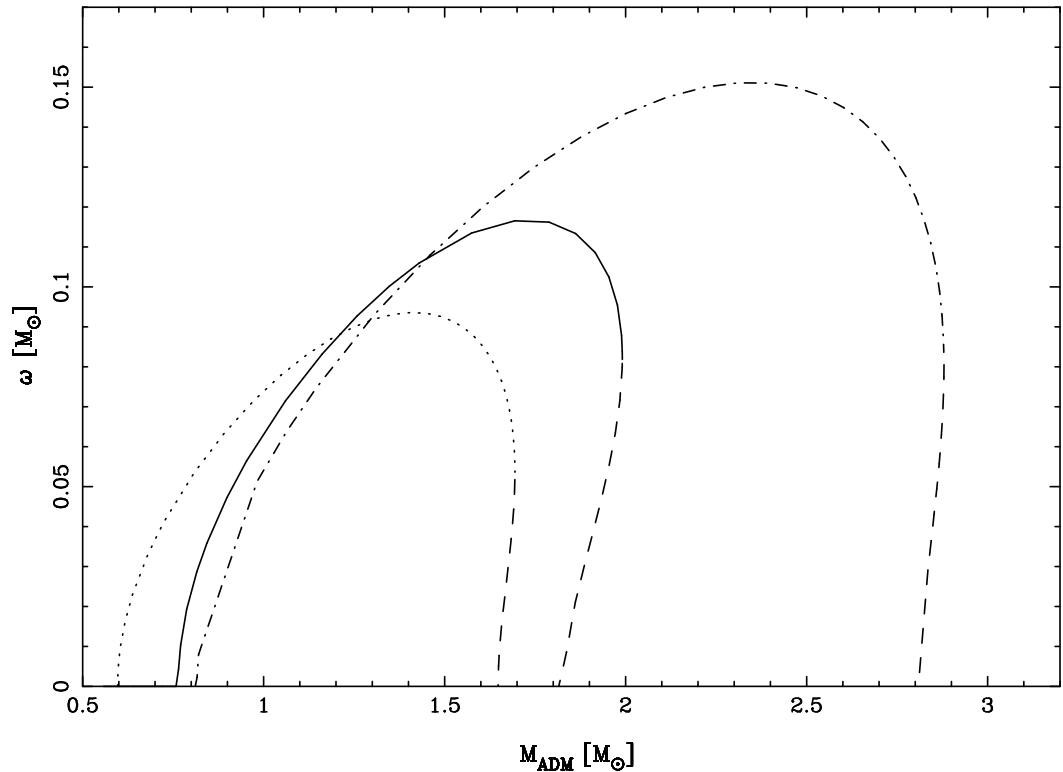} \hspace*{-2.5in} \caption[]{\label{f:ss}
Scalar charge vs $M_{\rm ADM}$ for the EOS labeled HKP (dash-dotted line),
DiazII (solid line), and PandN (dotted line) with $\xi=2$. The dashed
segments of the lines label the unstable configurations.} \end{figure*}
\vskip 1cm


\begin{figure*}
\vspace{1cm}
\psfig{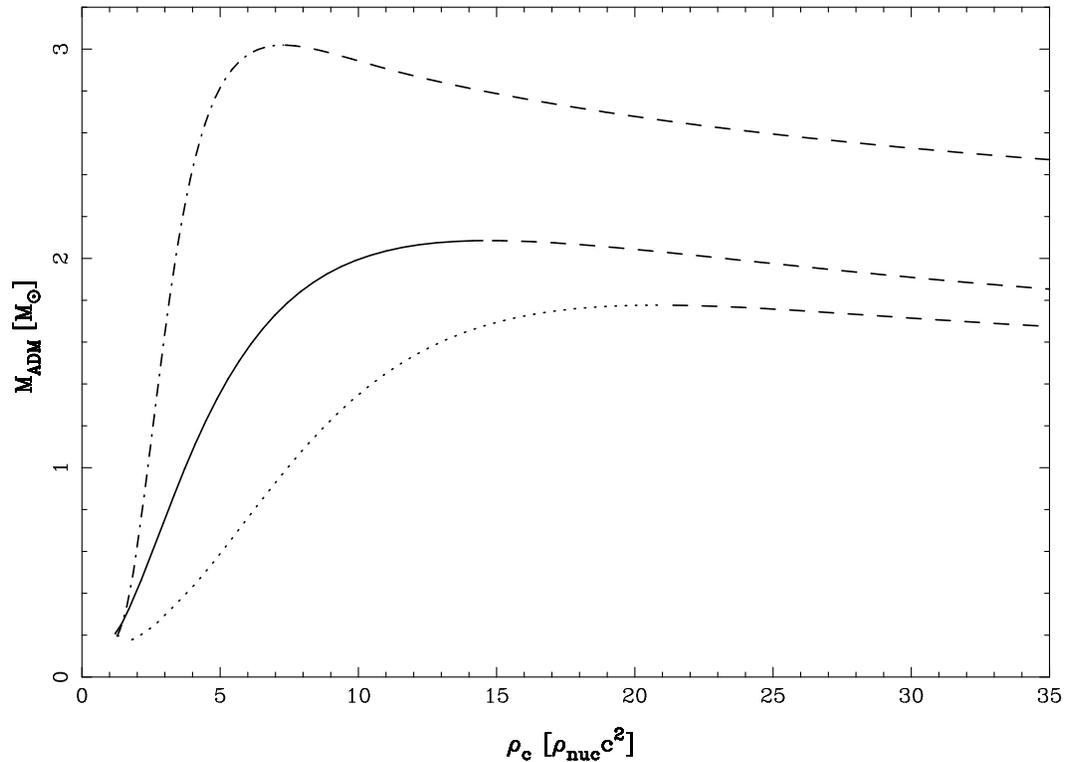}
\hspace*{-2.5in}
\caption[]{\label{f:mvsrho}
Total gravitational mass as a function of the central energy-density 
for the EOS HKP (dash-dotted line), DiazII (solid line) and 
PandN (dotted line) for $\xi=6$. The dashed segments of the lines correspond to 
unstable configurations.}
\end{figure*}


\end{document}